\documentclass[fleqn,10pt]{wlscirep}
\usepackage[utf8]{inputenc}
\usepackage[T1]{fontenc}
\title{ARCO-Mars: A Unified Cloud-Optimized Archive of Mars Atmosphere Reanalysis}

\author[1, * ]{Ananyo Bhattacharya}

\affil[1]{University of Michigan, Department of Climate and Space Sciences and Engineering, Ann Arbor, USA}

\affil[*]{ananyo@umich.edu}

\begin{abstract}
Long-term records of the Martian atmosphere based on general circulation models and reanalysis of atmospheric state variables are important to understand the diurnal, seasonal, and climatological changes of the planet. Atmospheric dynamics of the Martian atmosphere are strongly influenced by the characterization of dust lifting, solar insolation, and spatial variations in topography. We present ARCO-Mars, a unified Analysis-Ready Cloud-Optimized dataset providing integrated access to three independent Mars atmospheric reanalysis products: EMARS, MACDA, and OpenMARS spanning over Mars Years 24-35. These reanalyses assimilate thermal infrared retrievals from the MGS/TES, ODY/THEMIS, and MRO/MCS instruments, providing both two and three-dimensional surface and atmospheric state variables, including temperature, winds, surface pressure, and dust optical depth. The dataset is stored in Zarr v3 format and hosted on HuggingFace, enabling efficient cloud-based access without requiring local storage of the full archive. We compare the state variables between the three reanalysis products to identify systematic differences, attributed to differences in data assimilation and general circulation models. ARCO-Mars provides a community resource for Mars atmospheric science, numerical weather prediction validation, and machine learning applications, including weather forecasting and data assimilation.

\end{abstract}
\begin{document}

\flushbottom
\maketitle
%
%
\thispagestyle{empty}

\section*{Background and Summary}

Spacecraft missions to Mars have contributed to long-term records of Martian meteorology, including strong diurnal variations, dust storms, and chemical cycling of oxides of carbon and hydrogen. A comprehensive understanding of the general circulation motivated the development of atmospheric models with a wide range of complexity, from single-column radiative-convective equilibrium \cite{kasting1991co2, forget1997warming} to three-dimensional convection-allowing global-scale models \cite{basu2004simulation, richardson2007planetwrf}. The atmospheric dynamics and energy balance of Mars differ from that of Earth due to several key factors. It has a different radiative time scale compared to Earth. The presence of relatively tenuous surface pressure and the absence of water oceans influence the storage of heat and its transport. Spacecraft-based remote sensing and in-situ measurements also provide evidence of active hydrological and fluvial activity\cite{howard2005intense, kite2019geologic} in the past. Integration of spacecraft and telescopic observations with general circulation models constrains the natural cycles of water and carbon dioxide over the planet \cite{montmessin2004origin, montmessin2024mars}.\\

Data assimilation techniques have emerged as a crucial technological advancement to integrate physics-based forecasts with heterogeneous observations of Earth's atmosphere. It has advanced the state of Earth system modeling, examples include: satellite radiance assimilation for hurricane analysis \cite{liu2006radiance}, soil moisture assimilation in land-atmosphere modeling \cite{schaake2004intercomparison}, and radar reflectivity assimilation for mesoscale systems \cite{wang2013indirect}. Assimilation data products like ERA5\cite{hersbach2020era5}, MERRA2\cite{gelaro2017modern}, and JRA-55 \cite{kobayashi2015jra} provide an accessible record of spatio-temporal changes in regional and global climatology. In recent years, advancements in machine learning algorithms have leveraged reanalysis data products to train a large number of AI emulators for atmospheric dynamics \cite{weyn2021sub, lam2023learning}. Innovations in data management and metadata catalogs have increased the accessibility of weather and climate data for a wide range of applications, from scientific research to commercial ventures in risk assessment.\\

Unlike Earth, the Martian atmosphere has relatively sparse data collected over decades of observations from orbiters, rovers, and landers. Martian reanalysis involves assimilation of temperature and column dust opacity retrievals from infrared radiometry and spectroscopy in Martian orbit. The Mars Global Surveyor (MGS) Thermal Emission Spectrometer (TES) \cite{smith2001thermal}, Mars Odyssey (ODY) Thermal Emission Imaging System (THEMIS) \cite{christensen2004thermal}, and Mars Reconnaissance Orbiter (MRO) Mars Climate Sounder (MCS) \cite{mccleese2007mars} radiometer instruments serve as the main sources of atmospheric observations. There are four general climatology and reanalysis products of the Martian atmosphere: (i) Mars Climate Database (MCD) \cite{millour2015mars}, (ii) Ensemble Mars Atmosphere Reanalysis System (EMARS) \cite{greybush2012ensemble, greybush2019ensemble}, (iii) Mars Analysis Correction Data Assimilation (MACDA) \cite{montabone2014mars}, and (iv) Open access to Mars Assimilated Remote Soundings (OpenMARS) \cite{holmes2020openmars}. \\

EMARS, MACDA, and OpenMARS are publicly accessible, and the data files are stored in netCDF with metadata conventions compliant with weather and climate data. However, the analysis of Martian reanalysis data requires downloading and processing a large number of files, with each ranging from hundreds of MB to a few GB. A unified workflow and archive for storing the Mars reanalysis data provides accessibility for education and research purposes. It also benefits a systematic intercomparison of various general circulation models and assimilation methods. Martian reanalysis data also aid in the strategic development of engineering systems for Martian exploration, including observation system simulation experiments for mission planning and state estimation for entry, descent, and landing.\\

In this work, we present a unified database of analysis-ready Mars reanalysis data in a cloud-optimized format compliant with metadata conventions for weather and climate applications. The archival data from EMARS, MACDA, and OpenMars are processed and compressed into a publicly accessible format. The next sections describe the methods for data processing, verification of the new data product, and examples of systematic intercomparison of Mars reanalysis datasets.

\section*{Methods}

We use the Zarr data format for creating unified datasets for Mars surface and atmosphere reanalysis. Zarr is a widely used storage format for high-dimensional arrays. The processing of a Zarr file involves data chunk generation by dividing data into different subdomains. Chunks undergo data compression combined with metadata information necessary for locating and accessing variables. Metadata is stored in separate JavaScript Object Notation (JSON) files. It is compatible with Python libraries for data manipulation and is widely used in the geoscience and planetary science community. Zarr provides data processing capabilities on par with netCDF and hierarchical data format version 5 (HDF5), along with options for data compression and multiprocessing. In recent years, it has been developed for cloud-native applications in storing high-resolution numerical weather prediction, radar reflectivity \cite{gowan2022using}. Accessibility of cloud-compatible data formats enables efficient reading and writing of scientific data products, creating analysis-ready data products for wide-scale use \cite{abernathey2021cloud}. \\

We developed a Python-based data processing pipeline for EMARS, OpenMARS, and MACDA files. The netCDF files from each database are downloaded and stored in a private server. MACDA v2.0 is processed in bulk amounts directly using the UK Center for Environmental Data Analysis (CEDA) FTP service. The files are processed by MY and combined to generate a common Zarr file for each type of reanalysis product. The Blosc algorithm within the Python Zarr library is used for lossless data compression \cite{blosc}. A Python code is used to correct the metadata information to comply with Climate and Forecast (CF) metadata conventions \cite{eaton2020netcdf}, followed by a CF compliance check using the CEDA CF-checker tool. Post-completion of metadata correction and compliance check, the Zarr files are uploaded to Huggingface. The newly generated data products are renamed as ARCO-EMARS\cite{arco_emars}, ARCO-MACDA\cite{arco_macda}, and ARCO-OpenMARS\cite{arco_openmars}, respectively. Huggingface provides a platform to seamlessly query and read Mars data from cloud interface. Python-based fsspec package is used to access Zarr files on Huggingface server \cite{fsspec_docs}. \\

\begin{figure}[ht]
\centering
\includegraphics[width=\linewidth]{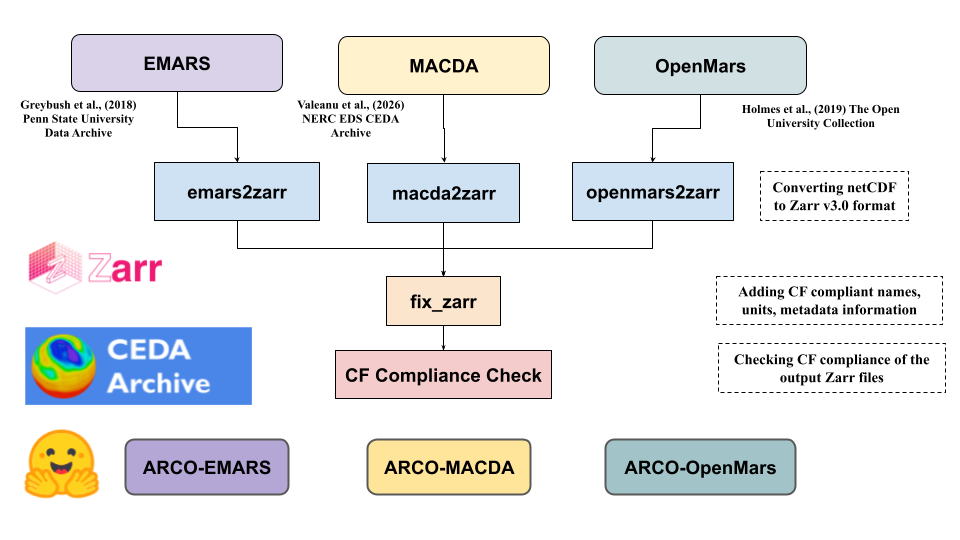}
\caption{Data processing pipeline to generate analysis-ready cloud-optimized (ARCO) Mars surface and atmosphere reanalysis data using archival data from EMARS, MACDA and OpenMARS reanalysis products}
\label{fig:fig1}
\end{figure}

\section*{Data Records}

We analyze the publicly available data products for EMARS, MACDA, and OpenMars. EMARS v1.0 is accessible through the Penn State University Data Commons \cite{greybush2019ensemble}. It uses the Geophysical Fluid Dynamics Laboratory (GFDL) Mars Global Climate Model (MGCM) \cite{john1996comprehensive} for the forecast model. EMARS follows a hybrid sigma‐pressure coordinate system with a terrain following sigma levels near the surface. The data assimilation component is an ensemble-based method based on the Local Ensemble Transform Kalman Filter (LETKF) \cite{greybush2012ensemble}. Both MGCM and EMARS have been applied to study planetary tides, transient eddies \cite{greybush2019transient}, water and dust cycles on Mars. The assimilation system uses retrieval temperature profiles from TES and MCS. Dust is represented by column dust opacity retrieved by the same sources. \\

Both OpenMARS\cite{holmes2020openmars} and MACDA\cite{montabone2014mars} reanalyses products are based on the UK spectral version of the Laboratoire de Météorologie Dynamique (LMD) MGCM \cite{forget1999improved}. The vertical coordinates are based on sigma levels defined by the ratio of atmospheric pressure to the surface pressure. MACDA v1.0 was the first surface and atmosphere reanalysis for Mars \cite{montabone2014mars}. As a part of the assimilation system, both MACDA and OpenMars use temperature and dust opacity retrievals from MCS and TES. The reanalysis products are generated using a sequential method, Analysis Correction (AC) scheme originally developed at UK Met-Office \cite{lorenc1991meteorological}. MACDA v2.0 also includes THEMIS retrieval to bridge observation gaps \cite{https://doi.org/10.5285/cd037a9ea387438fabf4d674dbe53088}. OpenMARS also considers TES total column water vapor \cite{steele2014seasonal}, and total column ozone \cite{perrier2006global} available from Mars Express (MEx) Spectrometer for the Investigation of the Characteristics of the Atmosphere of Mars (SPICAM). MACDA v2.0 is publicly available on the CEDA archive. OpenMARS is available on the Open University data portal.\\

\begin{figure}[ht]
\centering
\includegraphics[width=\linewidth]{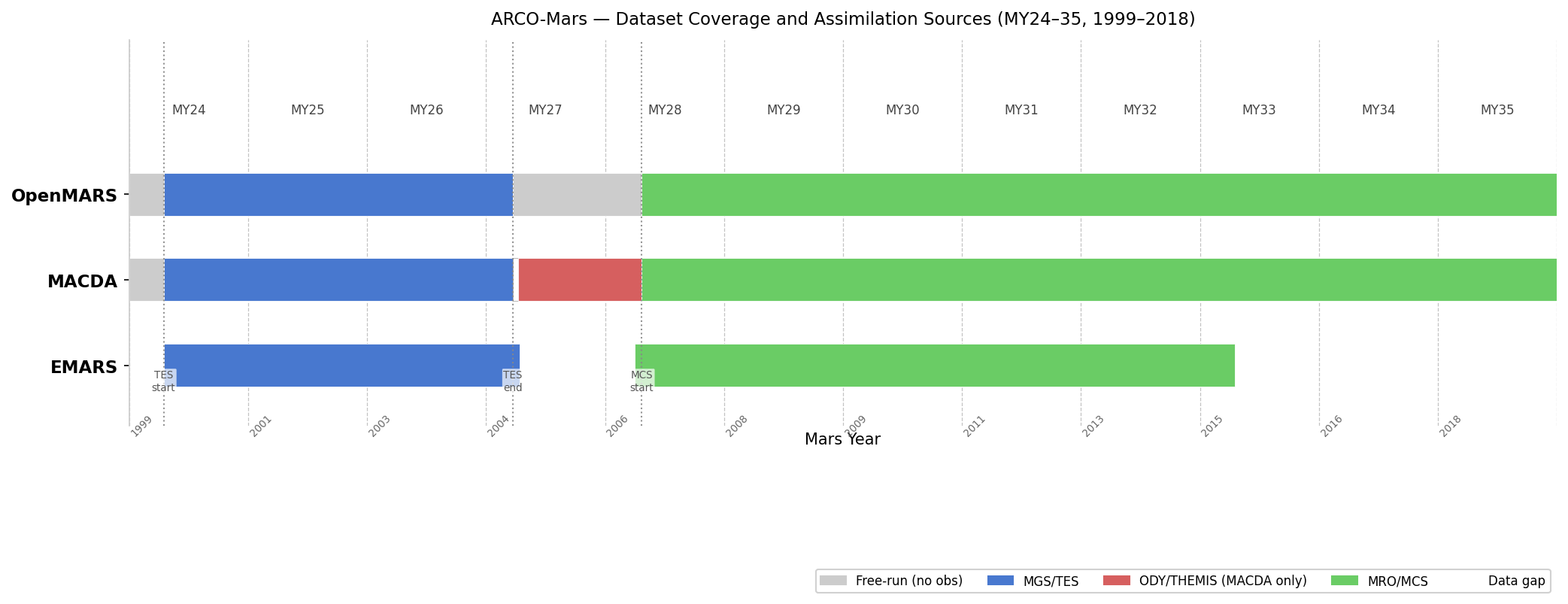}
\caption{Illustration of Mars reanalysis data, sources of observation data, and data gaps in EMARS, MACDA, and OpenMars. MGS/TES and MRO/MCS observations are common data to the assimilation scheme. MACDA v2.0 supplements the data gap with ODY/THEMIS observations, and OpenMARS provides the background run for regions with a lack of observation data for assimilation.}
\label{fig:fig2}
\end{figure}

For our analysis, we consider the analysis mean and spread data for EMARS spanning the Martian Years 24 to 33. EMARS has a data gap between MY 27-28 due to a lack of TES and MCS observations. In the case of MACDA and OpenMars, the archival data only provides information about the mean value of the estimated surface and atmospheric states spanning MY 24 to 35. The MY 27-28 gap is filled with assimilation of THEMIS data in MACDA, whereas OpenMars stores the numerical forecast in places of data gaps. The observation for each reanalysis product and associated data gaps are shown in Fig. \ref{fig:fig2}. \\

All the datasets contain information about atmospheric pressure, temperature, and horizontal wind speeds. MACDA and OpenMars also contain information about dust optical depth, and surface variables for CO$_{2}$ ice, pressure, and air temperature. MACDA also provides three-dimensional dust mass mixing data that aid in assessing dust activity during global and regional dust storms. Tables 1-4 provide a description of all variables stored in the Zarr stores. \\

\begin{table}[h]
\centering
\caption{Summary of ARCO-Mars datasets hosted on HuggingFace. Time steps reflect the total number of reanalysis snapshots after processing. Grid dimensions are given as latitude × longitude × vertical levels.}
\begin{tabular}{lllll}
\hline
\textbf{Dataset} & \textbf{MY} & \textbf{Grid (lat×lon×lev)} & 
\textbf{Assimilation} & \textbf{DOI} \\
\hline
ARCO-EMARS    & 24--33 & 36×60×28  & LETKF (TES, MCS)          & doi.org/10.57967/hf/8859 \\
ARCO-MACDA    & 24--35 & 36×72×35 & AC scheme (TES, THEMIS, MCS) & doi.org/10.57967/hf/8771 \\
ARCO-OpenMARS & 24--35 & 36×72×35  & AC scheme (TES, MCS, SPICAM)      &  doi.org/10.57967/hf/8741\\
\hline
\end{tabular}
\label{tab:datasets}
\end{table}

\begin{table}[ht]
\centering
\caption{\label{tab:variables}List of key variables in EMARS reanalysis.}
\begin{tabular}{|l|l|l|}
\hline
\textbf{Variable Name} & \textbf{Description} & \textbf{Units} \\
\hline
Surface\_geopotential & surface geopotential height & m$^2$/s$^2$ \\
\hline
T & Temperature & K \\
\hline
U & zonal wind & m/s \\
\hline
V & meridional wind & m/s \\
\hline
ps & surface pressure & Pa \\
\hline
\end{tabular}
\end{table}

\begin{table}[ht]
\centering
\caption{\label{tab:variables2}List of key variables in MACDA reanalysis.}
\begin{tabular}{|l|l|l|}
\hline
\textbf{Variable Name} & \textbf{Description} & \textbf{Units} \\
\hline
co2ice & surface frozen carbon dioxide amount & kg m$^{-2}$ \\
\hline
psurf & surface air pressure & Pa \\
\hline
tsurf & surface temperature & K \\
\hline
coldust & atmosphere optical thickness due to dust dry aerosol & 1 \\
\hline
swflux & surface downwelling shortwave flux in air & W m$^{-2}$ \\
\hline
lwflux & surface downwelling longwave flux in air & W m$^{-2}$ \\
\hline
temp & air temperature & K \\
\hline
uwind & eastward wind & m s$^{-1}$ \\
\hline
vwind & northward wind & m s$^{-1}$ \\
\hline
geop & geopotential & m$^2$ s$^{-2}$ \\
\hline
omega & lagrangian tendency of air pressure & Pa s$^{-1}$ \\
\hline
dustmmr & Dust mass mixing ratio & Kg Kg$^{-1}$ \\
\hline
\end{tabular}
\end{table}

\begin{table}[ht]
\centering
\caption{\label{tab:variables3}List of key variables in OpenMARS reanalysis.}
\begin{tabular}{|l|l|l|}
\hline
\textbf{Variable Name} & \textbf{Description} & \textbf{Units} \\
\hline
ps & Atmospheric pressure at the surface of Mars & Pa \\
\hline
tsurf & Atmospheric temperature at the surface of Mars & K \\
\hline
co2ice & Surface CO$_2$ ice & kg/m$^2$ \\
\hline
dustcol & Column dust optical depth & 1 \\
\hline
u & Zonal wind with negative/positive indicating easterly/westerly winds & m/s \\
\hline
v & Meridional wind with negative/positive indicating northerly/southerly winds & m/s \\
\hline
temp & The temperature measured at different levels of the Mars atmosphere & K \\
\hline
\end{tabular}
\end{table}

\section*{Technical Validation}

The ARCO-Mars datasets undergo verification through user-defined queries to the Huggingface platform. To assess surface and atmospheric variables, we analyze annual trends in surface pressure (P$_{s}$), surface air temperature (T$_{air}$), and dust optical depth ($\tau$) for EMARS, MACDA, and OpenMARS. We select MY 25, 28, and 32 as example years for intercomparison between reanalysis variables. Mars experienced a planet-encircling or global-scale dust storm during MY 28 that affected the temperature retrievals in the last few scale heights. The latitude weighted mean for each variable is expressed as a function of solar longitude (L$_{s}$). L$_{s}$ describes the seasonal cycle of Mars based on the Mars-Sun angle in various phases of Martian orbit. It starts at Northern Hemisphere Spring Equinox (L$_{s}$ = 0), and transcends into Northern Hemisphere Autumn Equinox (L$_{s}$ = 180). Regional and global dust storms (GDS) occur during the Southern Hemisphere Spring to Summer period (L$_{s}$ $\sim$ 180-360).\\

\begin{figure}[ht]
\centering
\includegraphics[width=\linewidth]{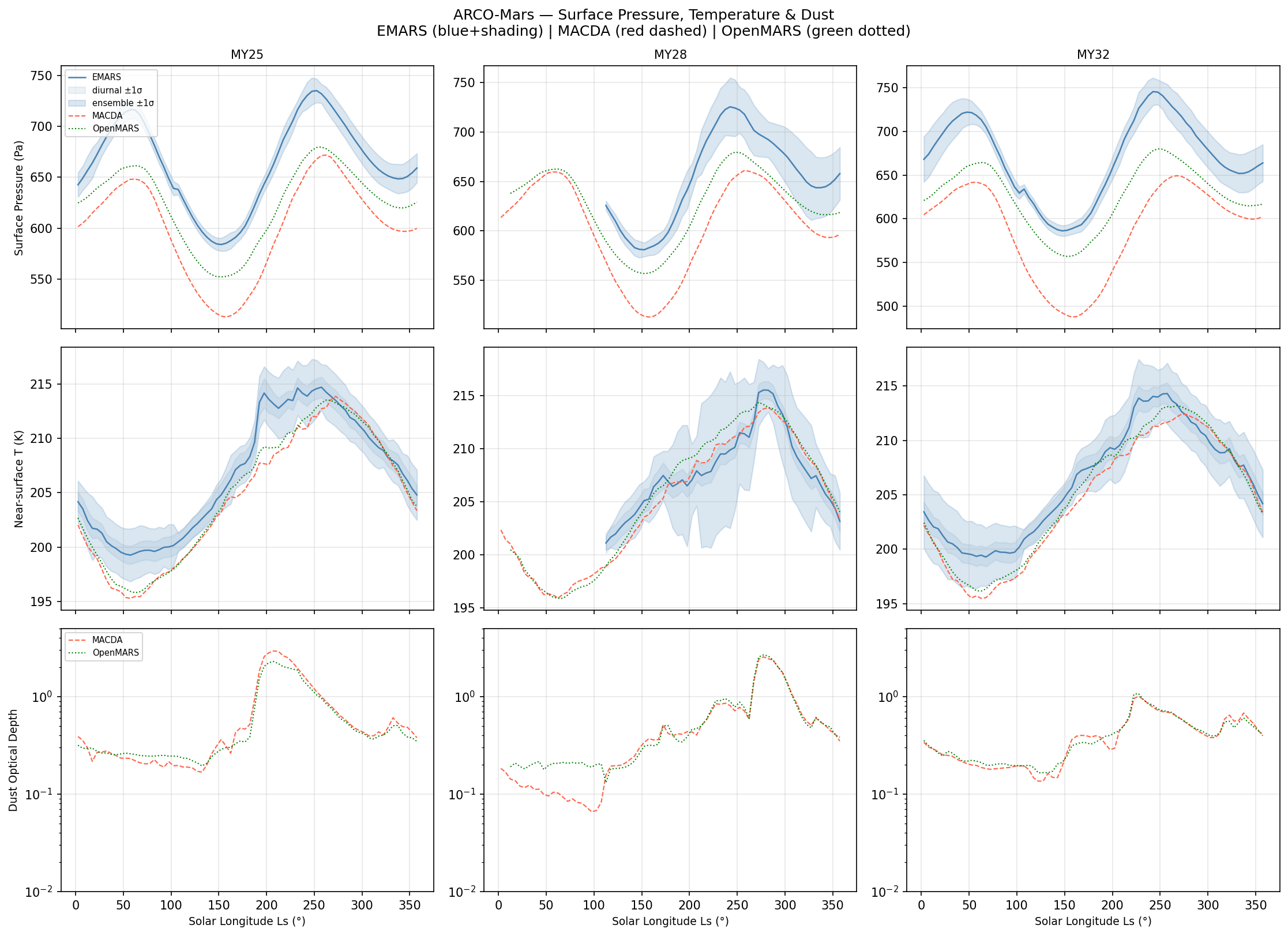}
\caption{Latitude weighted mean of surface pressure (Pa), near surface air temperature (K), and column dust optical depth (CDOD) at different Areocentric Solar Longitude (L$_{s}$). EMARS reanalysis data is expressed as a mean with uncertainty bounds from LETKF. EMARS surface variables show higher uncertainty during severe dust activity, coincident with high CDOD values. MACDA and OpenMARS CDOD values are normalized to 610 Pa surface pressure. MY 25, 28 and 32 are selected as representative cases of Martian atmospheric observations owing to different dust activity scenarios.}
\label{fig:fig3}
\end{figure}

Mean values of P$_{s}$ show a strong seasonal dependence as shown in Figure \ref{fig:fig3}. The peak-to-peak variation is found to range from 130 Pa. The reanalysis datasets show a consistent bias in P$_{s}$ values owing to differences in GCMs, and assimilation schemes. Both ARCO-MACDA and ARCO-OpenMARS show agreement in P$_{s}$ within a range of 10-20 Pa while the pressure level during equinox diverge by 30-50 Pa. It is to be noted that ARCO-EMARS maintains a significantly high value of mean P$_{s}$ throughout all seasons. The spread of analysis magnitude is observed to be higher during MY 28 due to large uncertainties in retrieval of atmospheric state during GDS. \\

Near-surface air temperature (T$_{s}$) values show a considerable amount of agreement between the three datasets, considering the uncertainties in EMARS reanalysis data. The T$_{s}$ uncertainty ranges from 5-10 K during MY 28 GDS, coincident with an elevated amount of column dust optical depth (CDOD) in both MACDA and OpenMARS. CDOD mean shows a large degree of agreement between the two datasets, even for regions with high dust activity. Differences in CDOD generally lie within a factor of two e.g. MY 25 and MY 28, Northern Spring to Summer period. Overall, these examples show the quality of assimilation and systematic differences in the treatment of surface-level quantities and dust. \\

The assimilation of temperature and dust opacity has implications on atmospheric thermal structure and wind speeds. Temperature structure and wind speeds are dynamically evolved in MGCMs. For EMARS, the ensemble forecast approach quantifies the background error covariance to adjust wind speeds in accordance with temperature assimilation. For all reanalysis datasets, wind speeds are largely found to be consistent with geostrophic balance. We extend the analysis by systematically understanding the differences in mean values of temperature ($\bar{T}$), zonal ($\bar{U}$), and meridional winds ($\bar{V}$) at 50 and 100 Pa. \\

\begin{figure}[ht]
\centering
\includegraphics[width=1\linewidth]{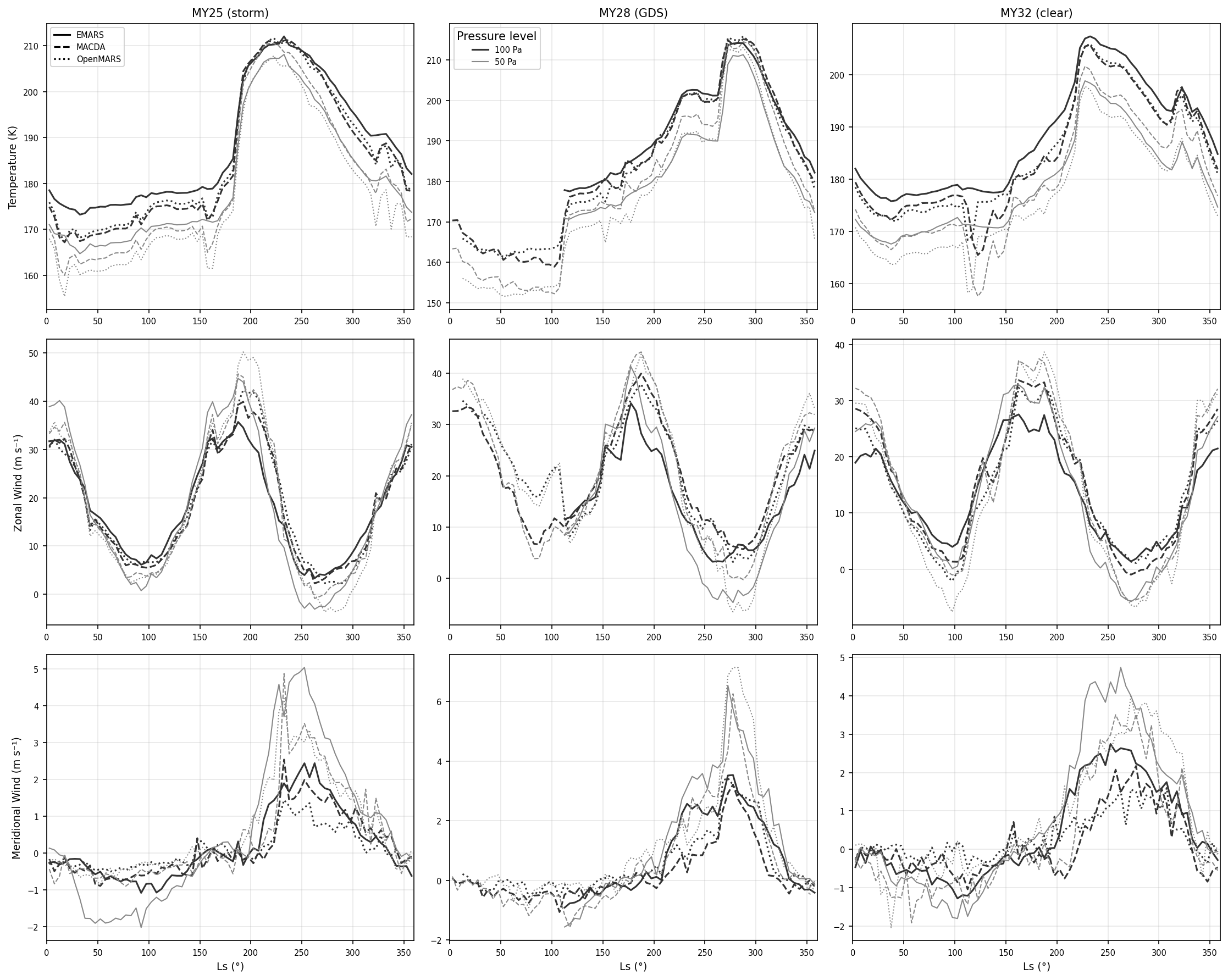}
\caption{Latitude weighted mean of gas temperature ($\bar{T}$), zonal ($\bar{U}$) and meridional wind speeds ($\bar{V}$) at two pressure levels: 50 and 100 Pa. MY 25, 28 and 32 are selected as representative cases of Martian atmospheric observations owing to different dust activity scenarios.}
\label{fig:fig4}
\end{figure}

We follow the same procedure to derive latitude-weighted mean values to look at global scale differences in vertical structure (Fig. \ref{fig:fig4}). Temperature generally decreases in the vertical direction with pressure level across all three datasets. However, the mean temperatures at 50 and 100 Pa converge at regions during Southern Spring and Summer, coincident with the onset of dust storm activity. The wind patterns show a strong influence of seasons for the representative MY. Peak-to-peak variations in $\bar{U}$ and $\bar{V}$ are found to be stronger in OpenMARS and MACDA. Low-level jets on Mars are biased in different MGCMs, leading to divergence in their mean values. Temperature assimilation also leads to relatively strong weakening of these jets in EMARS, consistent with our analysis.

\section*{Mars Global Dust Storm}

Here we show an example of atmospheric temperature and dust activity during the MY 28 GDS. \textit{fsspec} based query of the ARCO-Mars databases produces the latitudinal average of vertical temperature profiles during a high CDOD period. Equator (-10$^{o}$ to 10$^{o}$ deg.) and high-latitude Northern hemisphere (70$^{o}$ to 90$^{o}$ deg.) show large systematic differences in temperature characterized by a poleward decrease of temperatures in lower atmosphere. Similarly, a poleward increase of temperatures between 10 Pa to top of the atmosphere is observed. It is consistent with MGCM studies of polar warming in middle atmosphere \cite{deming1986polar, mcdunn2013characterization}. \\

MACDA dust mass mixing ratio ranges between 10$^{-4}$ to 10$^{-6}$ in regions of high dust activity. The local changes in dust mass mixing are found to be consistent with corresponding values of global CDOD. Dust activity is found to be largely concentrated in the Southern Hemisphere; however, the peak magnitude of CDOD lies around the equator and scattered parts of low-latitude hemispheres (-30$^{o}$ to 30$^{o}$ deg.). The high-latitude Northern Hemisphere shows the presence of multiple decks of dust clouds with an overall lower magnitude of CDOD. The presence of a middle atmosphere dust deck is found to be consistent with regions of polar warming.

\begin{figure}[ht]
\centering
\includegraphics[width=1\linewidth]{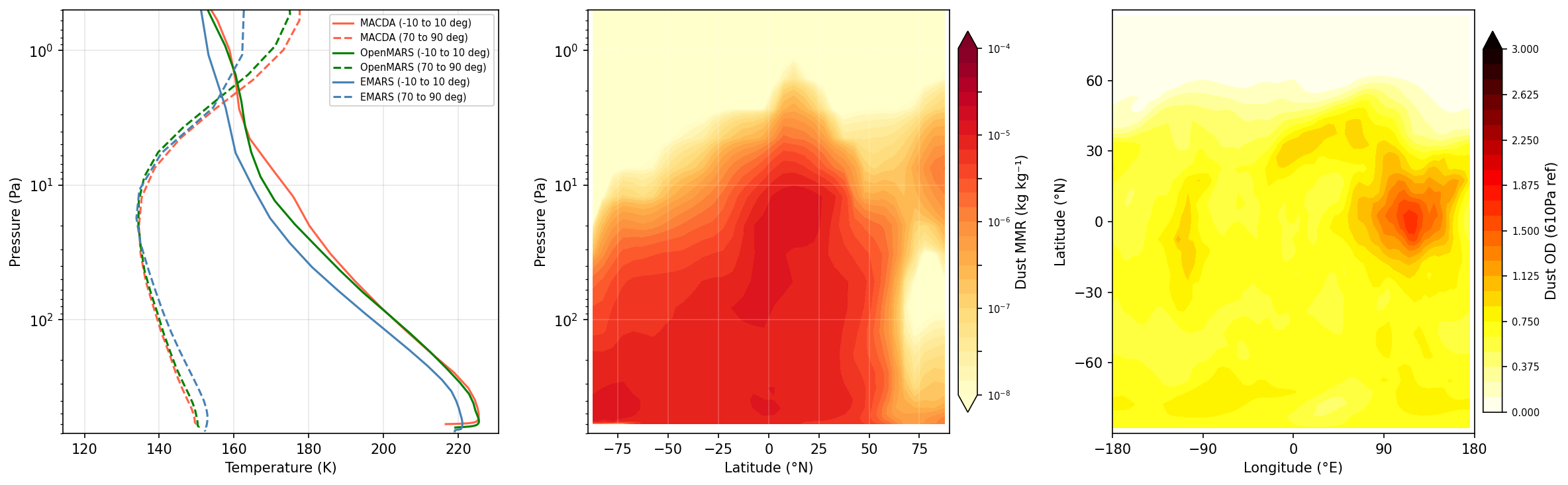}
\caption{First Panel: latitude-averaged temperature profiles of Martian atmosphere at equator and polar region during MY 28 GDS. Middle atmosphere temperature enhancement shows polar warming under intense dust activity. Second Panel: longitudinal average of dust mixing ratio profiles across all latitudes. Third Panel: CDOD corresponding to the dust mixing ratio profile in Second Panel.}
\label{fig:fig5}
\end{figure}

\section*{User Notes}

\subsection*{Example Code}

ARCO-Mars datasets are accessible directly from HuggingFace without local download using the \texttt{fsspec} and \texttt{xarray} Python libraries. The following code snippet demonstrates access to ARCO-MACDA and data manipulation for its variables:

\begin{verbatim}
import xarray as xr
import fsspec
import numpy as np


#Loading ARCO-MACDA from Huggingface Repository
fs    = fsspec.filesystem("hf")
store = fs.get_mapper("datasets/ananyo01/ARCO-MACDA/macda_combined.zarr")
ds    = xr.open_zarr(store, decode_times=False)


\end{verbatim}

Example code snippet to access the ARCO-EMARS reanalysis mean and spread values from Huggingface:

\begin{verbatim}
import xarray as xr
import fsspec
import numpy as np


#Loading ARCO-EMARS from Huggingface Repository
fs    = fsspec.filesystem("hf")

store_m = fs.get_mapper("datasets/ananyo01/ARCO-EMARS/emars_combined.zarr")
store_s = fs.get_mapper("datasets/ananyo01/ARCO-EMARS/emars_sprd.zarr")

de_m = xr.open_zarr(store_m, decode_times=False)
de_s = xr.open_zarr(store_s, decode_times=False)

\end{verbatim}

\subsection*{Variables and Diagnostics}

ARCO-Mars provides a selected number of variables as described in Tables 1-3. In addition to temperature and wind speeds, certain reanalysis products also provide additional variables that serve as a diagnostic of surface and atmospheric evolution. Both MACDA and OpenMARS contains surface density of CO$_{2}$ ice along with CDOD. MACDA also provides information about air pressure tendency, longwave, and shortwave downwelling radiation at the surface.

\section*{Data Availability Statement}
The ARCO Mars reanalysis datasets: ARCO-EMARS\cite{arco_emars}, ARCO-MACDA\cite{arco_macda} and ARCO-OpenMARS\cite{arco_openmars} are publicly available on Huggingface. Python examples for accessing the Hugging Face data have been made publicly available on GitHub \cite{bhattacharya2026arcomarsexamples}. The data processing pipeline can be made available upon request.

\section*{Acknowledgements}
The author acknowledges computing resources provided by the University of Michigan.

\section*{Author contributions statement}

A.B. conceived the idea, designed the data processing pipeline, performed the analysis, and wrote the manuscript.

\bibliography{sample}

\end{document}